\documentclass[final,5p,times,twocolumn]{elsarticle}
\usepackage{amsmath,amssymb,bm}
\usepackage{xcolor}
\usepackage[colorlinks=true,citecolor=blue,linkcolor=blue,urlcolor=blue]{hyperref}
\biboptions{numbers,sort&compress}
\newcommand{\dd}{\mathrm{d}}
\journal{Physics Letters B}

\usepackage{stfloats}
\usepackage{flushend}
\usepackage{placeins}
\usepackage{tikz}
\definecolor{lime}{HTML}{A6CE39}

\DeclareRobustCommand\orcidJonathan{{\href{https://orcid.org/0000-0001-9291-0893}{\orcidicon}}}
\DeclareRobustCommand\orcidEdson{{\href{https://orcid.org/0000-0001-9929-5977}{\orcidicon}}}

\makeatletter
\newcommand{\authorEmail}[2]{\begingroup\@eadauthor={#1}\ead{#2}\endgroup}
\makeatother
\begin{document}
\begin{frontmatter}
\title{Strong-deflection scaling in nearly symmetric thin-shell wormholes}
\author[ifce]{Jonathan A. Rebou\c{c}as\,\orcidJonathan\hspace{2mm}}
\authorEmail{Jonathan A. Rebou\c{c}as}{jalvesreboucas@ifce.edu.br}
\author[ufca]{Edson Otoniel\,\orcidEdson\hspace{2mm}}
\authorEmail{Edson Otoniel}{edson.otoniel@ufca.edu.br}
\address[ifce]{Instituto Federal de Educa\c{c}\~ao, Ci\^encia e Tecnologia do Cear\'a (IFCE), Iguatu, CE, Brazil}
\address[ufca]{Universidade Federal do Cariri (UFCA), Instituto de Forma\c{c}\~ao de Educadores - IFE, R. Oleg\'ario Emidio de Araujo S/N, Brejo Santo - CE, 63.260-000 - Brazil}
\begin{abstract}
The angular winding of a returning light ray can grow while the family of admissible impacts disappears. We characterize this competition near reflection symmetry in a static thin-shell wormhole joining Schwarzschild exteriors of slightly different masses. Exact shell-frame matching fixes the reflected interval, and a normal coordinate at each nondegenerate photon-sphere maximum gives a composite angular expansion with controlled error on compact interior channel intervals. The result remains valid when the throat distance divided by the mass contrast tends to zero, stays finite or diverges. For a throat distance proportional to a positive power of the fractional mass difference, the width exponent increases from one toward two and then saturates, while the logarithmic winding coefficient decreases from three toward two. Fixed-throat configurations retain their finite-distance amplitude and coefficient three. Reference-subtracted profiles connect the intermediate sector to finite and vanishing scaled throat positions. Independent quadratures test the scaling laws and resolve the slower convergence associated with changing scale ratios. The shrinking solid-angle aperture follows directly from the critical-impact gap. For an axial observer and a transparent test disk confined to the same exterior, the return-phase condition selects the admitted intersections; sampled near-symmetric configurations shift the first available global crossing to a later label. The closing aperture and increasing winding are therefore linked by the accessible radial extent, with disk access determined by the phase accumulated on the emitting segment.
\end{abstract}
\end{frontmatter}
\section{Introduction}\label{sec:introduction}
Images of M87* and Sagittarius A* have made the interpretation of strongly lensed emission a central problem in compact-object physics \cite{EHTM872019,EHTSgrA2022}. Comparisons between black holes and horizonless alternatives require separating the spacetime, the observer and the illuminating material \cite{CardosoPani2019,PerlickTsupko2022}. Photon surfaces characterize light trapping geometrically \cite{Claudel2001}, and exact lens equations describe image formation in static spherical metrics \cite{Perlick2004}. Near a nondegenerate photon sphere, the strong-deflection expansion has a logarithmic divergence \cite{Bozza2002,Tsukamoto2017}; the coefficient and accessible portions of a complete ray depend on the geometry.

Traversable wormholes provide hypothetical connections between distinct exterior regions \cite{MorrisThorne1988,BambiReview2021}. Their strong-lensing structure includes throat-dependent paths and relativistic image families \cite{Nandi2006,ShaikhWormholes2019,ShaikhUCO2019}. Other traversable-wormhole models include spin and torsion \cite{DiGrezia:2017daq} or quantum-corrected Casimir sources \cite{Battista:2024gud}. Thin-shell constructions localize the supporting stress on a junction \cite{Israel1966}, with surgically joined Schwarzschild exteriors furnishing a simple example \cite{Visser1989}. Asymmetric junctions and their radial stability have been studied separately \cite{Forghani2018,TsukamotoStability2021}. Here the junction is a prescribed static background for null propagation.

The optical interpretation additionally depends on source support. Early wormhole-shadow proposals considered millimeter interferometry \cite{Bambi2013}, while rotating geometries demonstrated the roles of spin and viewing inclination \cite{Nedkova2013}. Calculations with optically thin dust and thin disks illustrate how the material distribution and radiation from the opposite exterior affect an image \cite{Ohgami2015,Paul2020}. Geometrical ray access should therefore be established before assigning brightness to an additional branch. Complementary diagnostics use the Poynting--Robertson effect \cite{DeFalco:2020afv,DeFalco:2021klh} or reconstruct wormhole metrics from modified gravitational potentials \cite{DeFalco:2021ksd}.

For asymmetric thin shells, Wang et al. showed how energy matching maps the opposite photon sphere into an observer-side critical impact \cite{Wang}; related critical structures occur in joined Reissner--Nordstr\"om exteriors \cite{Wielgus2020}. Peng et al. identified crossing-and-return rays, their angular stages and their disk-transfer contributions in the Schwarzschild construction \cite{Peng2021}. Hayward and Bardeen profiles support related optical structures \cite{HaywardOptics2023,BardeenOptics2024}, and one- and two-disk images have been investigated for reflection-asymmetric wormholes in Palatini gravity \cite{Macedo2026}. Multiple-barrier image families in hairy black holes provide a comparison with a different interior and no junction \cite{Guo2022}.

Critical curves, photon subrings and emission-dependent brightness are distinct objects \cite{Gralla2019}. Black-hole subrings possess characteristic interferometric signatures \cite{Johnson2020}, and their shapes can retain geometrical information across source profiles \cite{GLM2020}. These results motivate studying the path structure of an additional wormhole channel, without transferring a brightness hierarchy to a different illumination model.

We examine the singular approach to reflection symmetry when the reflected impact interval closes while both Schwarzschild potential maxima remain nondegenerate. The throat's distance from the smaller photon sphere competes with the mass contrast. Combining a controlled expansion of the joint throat--mass limit with the separate fixed-throat analysis, we classify the width, accumulated angle and finite angular profiles. The mechanism differs from the nonlogarithmic divergence caused by a marginally unstable photon sphere \cite{Tsukamoto2020}. We also determine the shrinking solid-angle aperture and apply the established disk-crossing bookkeeping to the migration of returning intersections near symmetry.

This paper is organized as follows. Section~\ref{sec:geometry} fixes the geometry, Sec.~\ref{sec:scaling} derives the scaling laws, and Sec.~\ref{sec:disk_transfer} treats disk support. Section~\ref{sec:conclusions} presents our conclusions.

\section{Two barriers and a disappearing reflected channel}\label{sec:geometry}
Join two Schwarzschild exteriors at a static spherical throat of radius $a$. Their line elements are
\begin{equation}
 \dd s_i^2=-f_i(r)\dd t_i^2+\frac{\dd r^2}{f_i(r)}+r^2\dd\Omega^2,
 \qquad f_i(r)=1-\frac{2M_i}{r}.
 \label{eq:metric}
\end{equation}
The label $i=1,2$ denotes the exterior regions $\mathcal{M}_1$ and $\mathcal{M}_2$, $r$ is areal radius, $\dd\Omega^2$ the unit-sphere metric, and $t_i$ is normalized at its respective infinity. We use $G=c=1$ and signature $(-,+,+,+)$. Set $M_1=M$ and $M_2=M(1+\epsilon)$ with $M>0$ and dimensionless asymmetry $\epsilon>0$. The adopted range is $2M(1+\epsilon)<a<3M$, so the throat is timelike and both exterior photon spheres are retained.

The construction has the usual distributional supporting stress \cite{Israel1966,Visser1989}. With the Israel equation expressed in the shell orthonormal frame, the angular-curvature jump gives a surface density $-[\sqrt{f_1(a)}+\sqrt{f_2(a)}]/(4\pi a)$. It is negative throughout this static domain. We prescribe this background and study its geodesics; no material equation of state is required for the optical limit. Reflection symmetry is recovered at $\epsilon=0$, but the reflected channel considered below then has zero width.

The shell proper time satisfies $\dd\tau=\sqrt{f_i(a)}\,\dd t_i$. An equatorial photon has conserved segment energies $E_i=f_i\dd t_i/\dd\lambda$ and angular momentum $L=r^2\dd\phi/\dd\lambda$, where $\lambda$ is an affine parameter and $\phi$ is the azimuthal coordinate. Continuity of shell-frame frequency gives
\begin{equation}
 \frac{E_2}{E_1}=z,
 \qquad z=\sqrt{\frac{f_2(a)}{f_1(a)}}.
 \label{eq:matching}
\end{equation}
Here $z$ is the static energy-matching factor. Tangential phase continuity preserves $L$, so the two impact parameters $b_i=L/E_i$ obey $b_2=b_1/z$. We use $b=b_1$ to label the ray at the observing end and, by spherical symmetry, restrict attention to $b>0$. This distinction between the local matching and the two asymptotic impact parameters is essential to asymmetric-shell optics \cite{Wang}.

The null radial equation involves the potential $U_i(r)=f_i(r)/r^2$. Its maximum is at $r=3M_i$, with critical impact $3\sqrt3M_i$. Define $b_c=3\sqrt3M$. A ray arriving from end 1 passes its barrier if $b<b_c$. On side 2 it turns back toward the throat if $b_2>3\sqrt3M_2$. Hence the reflected interval is $b_-<b<b_+$, where
\begin{equation}
 b_+=b_c,
 \qquad b_-=b_c(1+\epsilon)
 \sqrt{\frac{a-2M(1+\epsilon)}{a-2M}}.
 \label{eq:edges}
\end{equation}
The lower edge is the contralateral critical impact expressed in end-1 units. Such crossing-and-return trajectories underlie additional image branches in earlier asymmetric-shell studies \cite{Wang,Peng2021}. Here their existence is an input to a more restricted question: how their entire impact interval and angular profile collapse as the masses approach equality.

Write $\Delta b=b_+-b_-$ and $\alpha=a/M$. To avoid subtracting nearly equal critical impacts, introduce the dimensionless exact expression
\begin{equation}
 D=\frac{\epsilon\left[2(3-\alpha)+(6-\alpha)\epsilon+2\epsilon^2\right]}{\alpha-2}.
 \label{eq:D}
\end{equation}
It equals $1-(b_-/b_c)^2$ and is positive in the stated two-barrier domain. The width is then
\begin{equation}
 \frac{\Delta b}{b_c}=\frac{D}{1+\sqrt{1-D}}.
 \label{eq:width}
\end{equation}
This form remains numerically well conditioned when the reflected channel becomes narrow. The denominator also makes clear that a small width is controlled by the difference of the barrier heights, rather than by a disappearance of either photon sphere separately.

Table~\ref{tab:critical_structure} summarizes the baseline used for the representative trajectories. Its two critical impacts delimit the reflected channel in the observer-side normalization.

\begin{table}[!tb]
\caption{Critical structure of the baseline configuration. Parameters are $M=M_1$, $M_2/M=1.02$ and $a/M=2.5$ ($\epsilon=0.02$). Paired radii are ordered by end 1 and end 2. The seed horizons are excised by the throat; both photon spheres are retained and unstable. Both critical impacts use the end-1 energy normalization. Numerical values are rounded independently.}
\label{tab:critical_structure}
\centering
\small
\renewcommand{\arraystretch}{1.15}
\begin{tabular}{lr}
\hline
Quantity & Value \\
\hline
Seed horizon radii, $r_{H,i}/M$ & $2,\ 2.04$ \\
Photon-sphere radii, $r_{\mathrm{ph},i}/M$ & $3,\ 3.06$ \\
Matching factor, $z$ & $0.959166$ \\
Lower critical impact, $b_-/M$ & $5.083654$ \\
Upper critical impact, $b_+/M$ & $5.196152$ \\
Reflected-channel width, $\Delta b/M$ & $0.112499$ \\
\hline
\end{tabular}
\end{table}

\subsection{Matched effective potentials}
The two optical barriers must be compared with the same energy normalization. With $E_1$ fixed, the relation in Eq.~\eqref{eq:matching} gives the potentials
\begin{equation}
 V_1(r)=U_1(r),\qquad V_2(r)=\frac{U_2(r)}{z^2}.
 \label{eq:matched_potential}
\end{equation}
The functions $V_i$ have dimensions of inverse length squared. A turning point in either exterior occurs at $V_i=1/b^2$, with $b$ always measured at end 1. The two potentials agree at the throat, since $z^2=f_2(a)/f_1(a)$. Their maxima are $1/b_+^2$ and $1/b_-^2$, respectively. The reflected channel therefore selects levels above the end-1 barrier and below the matched end-2 barrier.

For a single horizontal representation, define the dimensionless coordinate
\begin{equation}
 \chi=\begin{cases}(r-a)/M,&\text{end 1},\\-(r-a)/M,&\text{end 2}.\end{cases}
 \label{eq:glued_coordinate}
\end{equation}
This is a signed areal-radius displacement, not a proper distance. It joins the retained potentials at $\chi=0$ without identifying the two asymptotic clocks.

\begin{figure*}[!tb]
\begin{minipage}{\textwidth}
\centering\includegraphics[width=\textwidth]{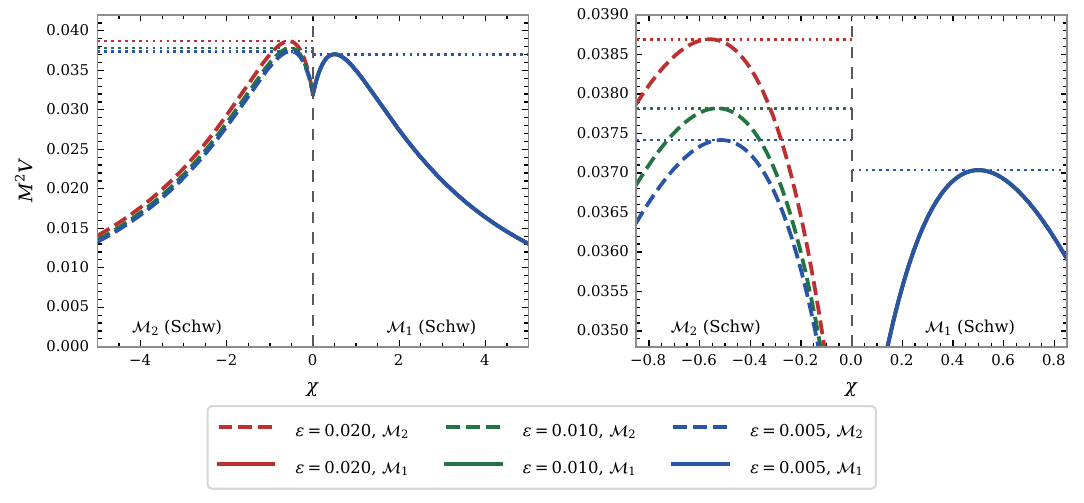}
\caption{Matched optical potentials at fixed $a/M=2.5$ for $\epsilon=0.02,0.01,0.005$. Schw denotes Schwarzschild. Color denotes the mass asymmetry; end 1 is solid and end 2 dashed. Dotted horizontal segments mark the two critical levels, and the vertical line marks the throat. Left: the retained geometry over $-5\leq\chi\leq5$. Right: a magnified interval around the two maxima; the lower throat value lies below this panel's vertical window. The ordinate is $M^2V_i$ in both panels. The end-1 curves and their critical levels coincide exactly for all three asymmetries.}
\label{fig:potential}
\end{minipage}
\end{figure*}
The two barriers in Fig.~\ref{fig:potential} retain distinct maxima even when their heights become close. The end-1 potential stays fixed because both $M$ and $a$ are fixed. The matched end-2 maximum approaches it from above as the mass asymmetry decreases. The range of horizontal ray levels between these maxima consequently contracts. Through Eq.~\eqref{eq:edges}, this is the same closing interval that appears in impact space; a larger allowed energy level corresponds to a smaller impact parameter.

The throat remains below both maxima in the configurations shown. A returning ray can therefore pass through the throat before reaching its end-2 turning point. Closing the critical-level gap does not require either Schwarzschild maximum to flatten or the throat to become a horizon. The local curvature of each barrier remains nonzero, a distinction that becomes relevant to the logarithmic limits derived below.

\subsection{Representative ray classes}
The same critical impacts divide incoming rays into three open classes. For $b>b_+$, the ray turns in end 1 before reaching the throat. For $b_-<b<b_+$, it crosses, turns in end 2 and returns to end 1. For $0<b<b_-$, it crosses and escapes through end 2. Equalities correspond to limiting critical rays and are excluded from the numerical trajectories.

To display the paths, use planar polar projections in each exterior,
\begin{equation}
 X=r\cos\phi,\qquad Y=r\sin\phi.
 \label{eq:ray_projection}
\end{equation}
The projection coordinates $X$ and $Y$ have dimensions of length. The displayed curves superpose these projections, with the exterior and direction identified by color. They do not represent an embedding of both exterior spatial geometries into one Euclidean plane.

\begin{figure*}[!tp]
\begin{minipage}{\textwidth}
\centering\includegraphics[width=\textwidth]{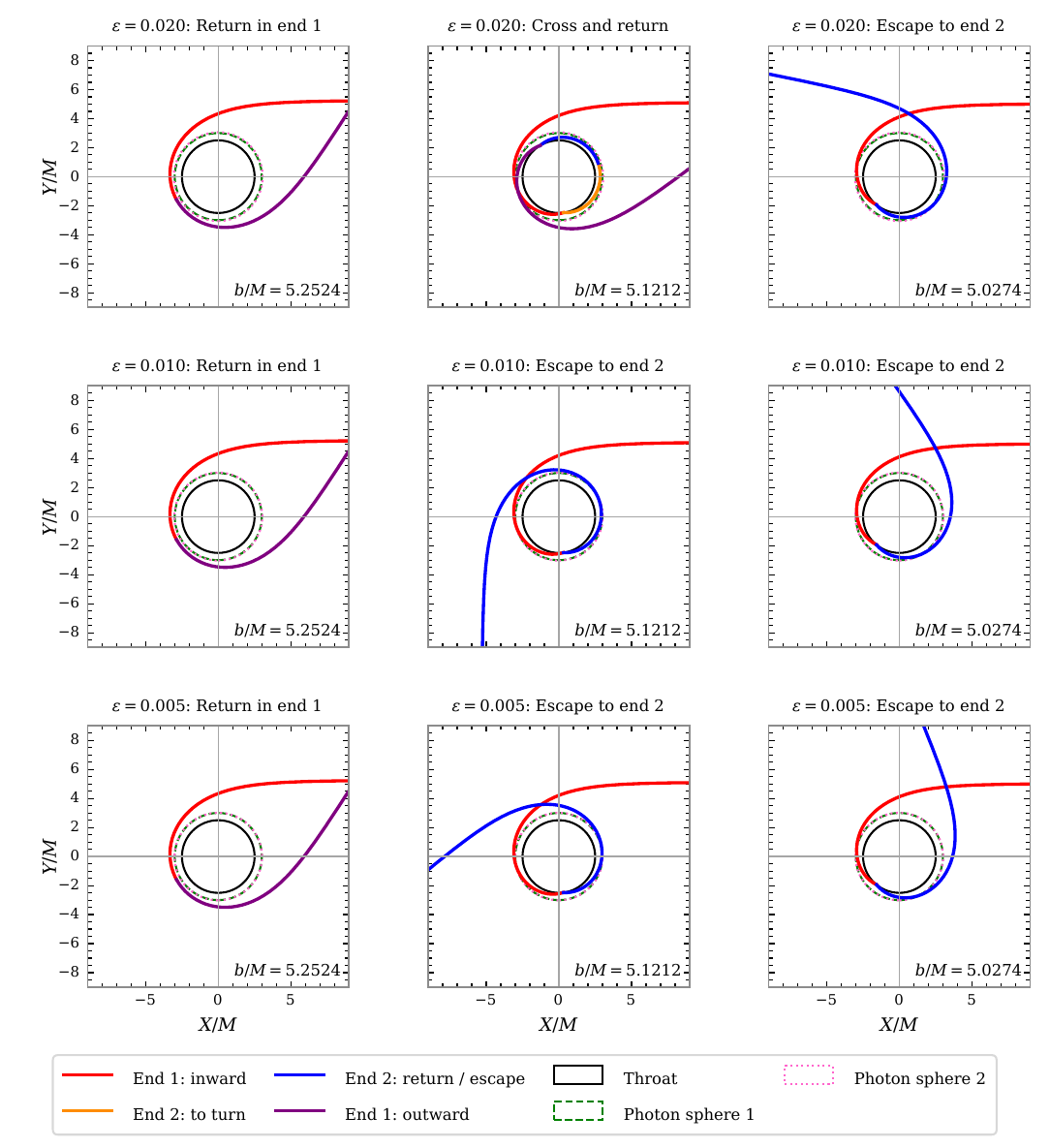}
\caption{Null trajectories at fixed $a/M=2.5$. Rows have $\epsilon=0.020$, $0.010$ and $0.005$, from top to bottom. The column impacts $b/M\simeq5.2524$, $5.1212$ and $5.0274$ represent the three ray classes at $\epsilon=0.020$ and are held fixed across rows. Every panel uses $-9\leq X/M,Y/M\leq9$ and equal spatial scales. The black circle is the throat; green and pink circles mark the photon-sphere radii $3M$ and $3M(1+\epsilon)$. Directional colors distinguish regional segments. On end 2, orange marks motion from the throat toward a turning point; blue marks the subsequent return to the throat or transmission to infinity. Only the upper central ray returns through the throat; the lower central rays escape into end 2. The finite display shows central portions of trajectories extending to their respective infinities.}
\label{fig:trajectories}
\end{minipage}
\end{figure*}
Figure~\ref{fig:trajectories} compares the fate of the same observer-side impacts as the contralateral mass changes. The upper row contains the three classes: ordinary end-1 scattering, crossing followed by reflection in end 2, and transmission. In the central column, the fixed impact is inside the reflected channel only at $\epsilon=0.020$. The lower edge $b_-/M$ rises from approximately $5.08365$ to $5.14208$ and $5.16965$ as $\epsilon$ decreases to $0.010$ and $0.005$, while $b_+/M$ stays at $5.19615$. Equation~\eqref{eq:edges} therefore places the same central impact below the lower edge in both later rows: the photon crosses the end-2 barrier and escapes instead of returning through the shell. These panels resolve a change of scattering class caused by the closing impact interval.

The first column provides an observer-side control. Its impact exceeds $b_+$, so none of these rays reaches the throat; their trajectories coincide because $M$ and $a$ are unchanged. The end-1 incoming segment likewise coincides across rows in each of the other columns. Their end-2 continuations depend on both the contralateral potential and the mapped impact $b_2=b/z$. In the right-hand column all three rays remain transmitted, although their projected escape directions vary. Equal impacts in this figure compare distinct static configurations; they do not describe the evolution of a photon. The normalized channel coordinate defined in Eq.~\eqref{eq:x} therefore varies between rows.

All displayed segments remain at $r\geq a$. The empty region inside the throat circle is excised, not a captured-ray domain, and the planar curves are regional projections rather than a spatial embedding. The nine numerical paths were checked against independent quadratures of the Schwarzschild radial cubic, including the omitted asymptotic angular tails; the largest complete-angle difference was below $2\times10^{-5}$ radians.

\section{Competing near-symmetry scales}\label{sec:scaling}
A ray remains within the shrinking reflected interval when its dimensionless relative impact
\begin{equation}
 x=\frac{b-b_-}{\Delta b},\qquad 0<x<1,
 \label{eq:x}
\end{equation}
is held fixed. A generic fixed physical impact instead changes ray class, as in Fig.~\ref{fig:trajectories}. Let $r_t$ be the inner end-2 turning point, $a<r_t<3M_2$. The complete accumulated azimuth is
\begin{equation}
 \begin{split}
 \Phi={}&2\int_a^\infty\frac{b\,\dd r}{r^2\sqrt{1-b^2U_1(r)}}\\
 &+2\int_a^{r_t}\frac{b_2\,\dd r}{r^2\sqrt{1-b_2^2U_2(r)}}.
 \end{split}\label{eq:angle}
\end{equation}
This dimensionless angle includes both end-1 passages and the full end-2 excursion; no straight-line reference angle is subtracted. Denote the one-way end-1 integral by $\varphi_1$ and the full end-2 term by $\varphi_2$, so $\Phi=2\varphi_1+\varphi_2$. The returning ray reaches the throat again and finally escapes at the angular stages \cite{Peng2021}
\begin{equation}
 n_2=\frac{\varphi_1+\varphi_2}{2\pi},\qquad
 n_3=\frac{\Phi}{2\pi}.
 \label{eq:orbit_stages}
\end{equation}
These counts distinguish segments of a single trajectory and will set the available disk-crossing phases.

\subsection{Width and approach paths}
Introduce the dimensionless throat distance $\delta=3-\alpha$. The adopted domain becomes $\delta>0$ and $\delta+2\epsilon<1$. We consider the power-law paths
\begin{equation}
 \delta=\kappa\epsilon^p,\qquad K=\frac{\delta}{\epsilon},
 \label{eq:paths}
\end{equation}
where $p>0$ and $\kappa>0$ are fixed dimensionless parameters, and $K$ compares the throat distance with the mass contrast. The admissibility condition is $\kappa\epsilon^p+2\epsilon<1$; it is satisfied at sufficiently small positive $\epsilon$. Every point is a distinct static configuration. A fixed throat is treated separately, or as $p=0$ with $0<\kappa<1$.

Writing $w=\Delta b/b_c$, expansion of Eqs.~\eqref{eq:D} and \eqref{eq:width} in the joint limit gives
\begin{equation}
 w=\left(\epsilon\delta+\frac32\epsilon^2\right)
 \left[1+O(\delta+\epsilon)\right].
 \label{eq:composite_width}
\end{equation}
The error is relative and independent of whether $K$ becomes small or large. Indeed, the exact numerator of $D$ is $2\epsilon\delta+3\epsilon^2+\delta\epsilon^2+2\epsilon^3$, and its denominator is $1-\delta$. The two displayed leading terms exchange dominance at $\delta\sim\epsilon$. Thus $0<p<1$ gives $w\sim\kappa\epsilon^{1+p}$, $p=1$ gives $w\sim(\kappa+3/2)\epsilon^2$, and $p>1$ gives $w\sim3\epsilon^2/2$. For a fixed $2<\alpha<3$, the exact expression instead yields $w=[(3-\alpha)/(\alpha-2)]\epsilon+O(\epsilon^2)$. Expanding in small $\delta$ is unnecessary in that case and would lose its finite-throat coefficient.

\begin{figure*}[!tb]
\begin{minipage}{\textwidth}
\centering\includegraphics[width=\textwidth]{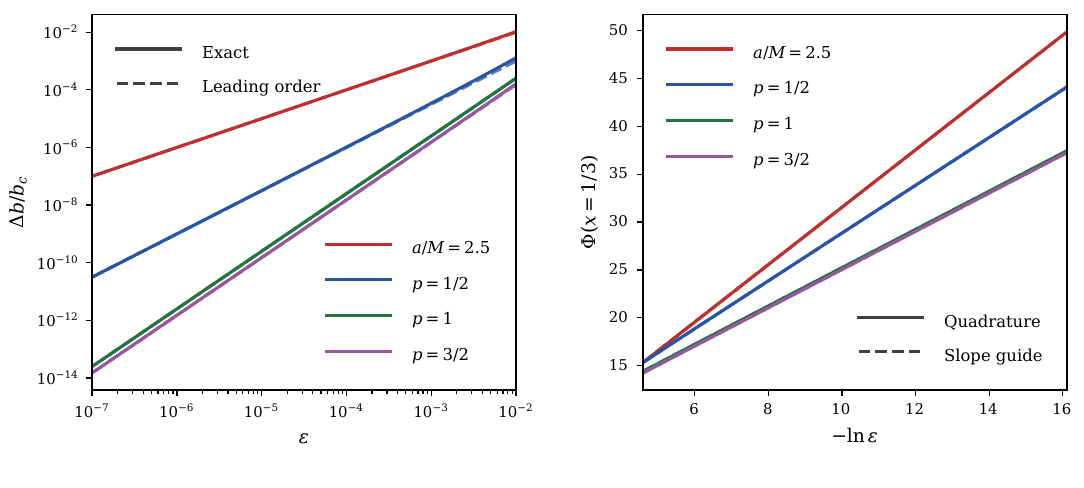}
\caption{Width and accumulated angle for fixed $a/M=2.5$ and for $a/M=3-\epsilon^p$, with $p=1/2,1,3/2$. All rays have $x=1/3$ and $M=1$. Left: exact dimensionless widths (solid) and leading-order widths from Table~\ref{tab:regimes} (dashed). Right: full angular integrals (solid); dashed guides have the analytic slopes $3,5/2,2,2$ and are aligned to the numerical value at $\epsilon=10^{-7}$. Their intercepts are not independent constant predictions. The common sweep is $10^{-7}\leq\epsilon\leq10^{-2}$.}
\label{fig:first}
\end{minipage}
\end{figure*}

\subsection{A composite accumulated angle}
The logarithmic mechanism can be controlled without assuming that $K$ remains fixed. For each exterior set $R_i=r/M_i$ and $\beta_i=b_i/(3\sqrt3M_i)$. Near its photon sphere, the signed dimensionless normal coordinate
\begin{equation}
 v_i=(R_i-3)\sqrt{\frac{3(R_i+6)}{R_i^3}}
 \label{eq:normal_coordinate}
\end{equation}
obeys $27M_i^2U_i=1-v_i^2/3$ exactly. It is monotonic near $R_i=3$, with $v_i=R_i-3+O((R_i-3)^2)$. Define the positive critical radial scales
\begin{equation}
 \rho_+^2=3(\beta_1^{-2}-1),\qquad
 \rho_-^2=3(1-\beta_2^{-2}).
 \label{eq:radial_scales}
\end{equation}
They satisfy $\rho_+^2=6w(1-x)[1+O(w)]$ and $\rho_-^2=6wx[1+O(w)]$. In the normal coordinate, the exact angular weights are $J(v_i)/\sqrt{v_i^2+\rho_+^2}$ on end 1 and $J(v_i)/\sqrt{v_i^2-\rho_-^2}$ on end 2, where $J(v_i)=9R_i^{-2}\dd R_i/\dd v_i=1+O(v_i)$. The inner turning point is $v_2=-\rho_-$.

To match the end-1 local integral to the outer ray, take a dimensionless separation $\ell=\sqrt{\delta+\epsilon}$. Both the throat distance and the critical radial scales are smaller than $\ell$. Integrating the leading weight gives $\operatorname{arsinh}(\ell/\rho_+)+\operatorname{arsinh}(\delta/\rho_+)$. The critical outer integral contributes $-\ln\ell+\ln[18(2-\sqrt3)]+O(\ell)$, which follows directly by the substitution $\sqrt{1+6/R_1}$. The artificial separation cancels. On end 2, the leading integral from the throat to the turning point is an inverse hyperbolic cosine. Hence
\begin{equation}
 \begin{split}
 \varphi_1={}&-\ln\rho_++\operatorname{arsinh}\frac{\delta}{\rho_+}
 +\frac{C_0}{2}+O\!\left(\sqrt{\delta+\epsilon}\right),\\
 \varphi_2={}&2\operatorname{arcosh}\frac{\delta+3\epsilon}{\rho_-}
 +O(\delta+\epsilon),
 \end{split}\label{eq:local_angles}
\end{equation}
where $C_0=2\ln[36(2-\sqrt3)]$ is dimensionless. The factor $1+\epsilon$ in the end-2 radial normalization changes the second line only at the stated error order.

The remainder control uses the exact normal-coordinate weights. Replacing $J(v)$ by one gives an error bounded by $O(\ell+\delta)$ on end 1 and by $O(\delta+\epsilon)$ on the turning segment, since the relevant integrals of $|v|/\sqrt{v^2\pm\rho^2}$ are finite. Replacing the outer end-1 ray by its critical value costs $O(w/\ell^2)=O(\delta+\epsilon)$. Endpoint replacements remain bounded for $x$ in any fixed $[x_*,1-x_*]$, $0<x_*<1/2$: the leading end-2 inverse-hyperbolic argument has square $[(K+3)^2/(6K+9)]/x\geq1/x$. This prevents it from approaching one in that interval, independently of $K\geq0$.

Define $d_K=K+3/2$ and the dimensionless crossover function
\begin{equation}
 \begin{split}
 G_K(x)={}&-\ln[6d_K(1-x)]\\
 &+2\operatorname{arsinh}\frac{K}{\sqrt{6d_K(1-x)}}\\
 &+2\operatorname{arcosh}\frac{K+3}{\sqrt{6d_Kx}}.
 \end{split}\label{eq:crossover_shape}
\end{equation}
Using Eq.~\eqref{eq:composite_width} in Eq.~\eqref{eq:local_angles} yields the composite expansion
\begin{equation}
 \Phi=-2\ln\epsilon+C_0+G_K(x)
 +O\!\left(\sqrt{\delta+\epsilon}\right).
 \label{eq:composite_angle}
\end{equation}
The estimate is uniform on each such compact channel interval as $\delta,\epsilon\to0$, including varying positive $K$. It does not include fixed positive $\delta$ or simultaneous approach to either critical edge.

\begin{figure*}[!tb]
\begin{minipage}{\textwidth}
\centering\includegraphics[width=\textwidth]{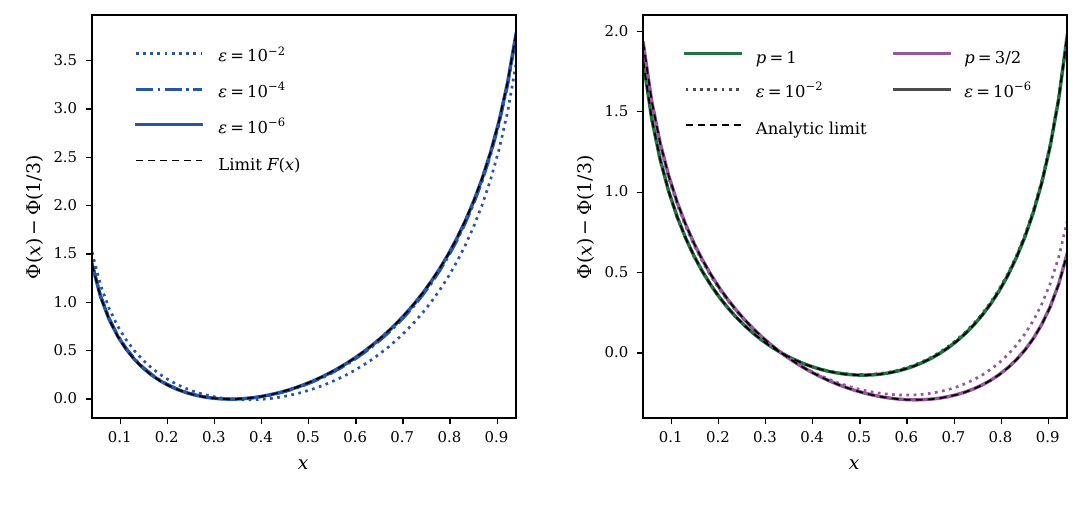}
\caption{Reference-subtracted angular profiles for $\kappa=1$ on $0.04\leq x\leq0.94$. Left: $p=1/2$ at $\epsilon=10^{-2},10^{-4},10^{-6}$. Right: $p=1$ (green) and $p=3/2$ (purple), each at $\epsilon=10^{-2},10^{-6}$; color identifies $p$ and line style identifies $\epsilon$. Black dashed curves denote $F(x)$ on the left, and $G_1(x)-G_1(1/3)$ and $G_0(x)-G_0(1/3)$ for the respective families on the right. The solid $\epsilon=10^{-6}$ curves nearly coincide with these analytic limits. The vertical scales differ; angles are in radians.}
\label{fig:profiles}
\end{minipage}
\end{figure*}

\subsection{Width and winding exponents}
For $K\to\infty$, expansion of Eq.~\eqref{eq:crossover_shape} gives $G_K=\ln K+\ln(2/27)-\ln x-2\ln(1-x)+O(K^{-1})$. When $0<p<1$, $K=\kappa\epsilon^{p-1}$ diverges, so
\begin{equation}
 \begin{split}
 \Phi={}&-(3-p)\ln\epsilon+C_0+\ln\frac{2\kappa}{27}\\
 &-\ln x-2\ln(1-x)+o(1).
 \end{split}\label{eq:intermediate_angle}
\end{equation}
The end-1 contribution $2\varphi_1$ has coefficient two, while $\varphi_2$ contributes $1-p$. In critical radial units, the throat remains far from the photon-sphere core on its inner side, even though its absolute distance from the maximum tends to zero. Increasing $p$ shortens the logarithmically accessible part of that excursion.

At $p=1$, $K=\kappa$ is finite and Eq.~\eqref{eq:composite_angle} gives coefficient two with shape $G_\kappa$. At $p>1$, $K\to0$, giving the same coefficient with shape $G_0$. Both cases have a bounded end-2 excursion in critical radial units. For a fixed throat, the two complete end-1 crossings and the end-2 turning excursion instead give
\begin{equation}
 \Phi=-3\ln\epsilon+C(\alpha)-\ln x-2\ln(1-x)+o(1),
 \label{eq:fixed_profile}
\end{equation}
where the finite constant $C(\alpha)$ depends on the fixed throat. This uses the local nondegenerate-maximum expansion underlying strong deflection \cite{Bozza2002,Tsukamoto2017}, with the full integration segments retained.

Let $q$ be the leading width exponent and $\gamma$ the coefficient of $-\ln\epsilon$ in $\Phi$. Including the fixed-throat case as $p=0$, the classification is
\begin{equation}
 q=1+\min(p,1),\qquad \gamma=3-\min(p,1).
 \label{eq:exponents}
\end{equation}
Their sum is four within this model and this class of paths, using the physical fractional mass contrast as small parameter. Table~\ref{tab:regimes} separates the exponents from the amplitudes. The change at $p=1$ comes from the accessible radial extent, while both Schwarzschild maxima keep nonzero curvature.

\begin{table*}[!tb]
\caption{Near-symmetry regimes at fixed interior $x$. The positive leading width amplitude $A$ is defined by $w\sim A\epsilon^q$; the accumulated angle is $\Phi\sim-\gamma\ln\epsilon$. The final column gives the limiting reference-subtracted profile, with $x_0=1/3$. Power-law paths have fixed $\kappa>0$; the fixed-throat row retains its exact finite-$\alpha$ amplitude.}
\label{tab:regimes}
\centering
\begin{tabular}{lcccc}
\hline
Approach & $A$ & $q$ & $\gamma$ & Finite profile \\
\hline
Fixed $2<\alpha<3$ & $(3-\alpha)/(\alpha-2)$ & $1$ & $3$ & $F(x)$ \\
$0<p<1$ & $\kappa$ & $1+p$ & $3-p$ & $F(x)$ \\
$p=1$ & $\kappa+3/2$ & $2$ & $2$ & $G_\kappa(x)-G_\kappa(x_0)$ \\
$p>1$ & $3/2$ & $2$ & $2$ & $G_0(x)-G_0(x_0)$ \\
\hline
\end{tabular}
\end{table*}

The intermediate path in Fig.~\ref{fig:first} fills the gap between the fixed-throat and finite-$K$ scalings. The $p=1$ and $p=3/2$ widths approach the same exponent with different amplitudes, while their angles approach equal slopes with different finite parts. Over the final adjacent sampling interval ending at $\epsilon=10^{-7}$, the numerical width exponents and angular slopes differ from Table~\ref{tab:regimes} by less than $5\times10^{-4}$ and $10^{-4}$, respectively, across the four paths. Convergence can be slow when $p$ is close to zero or one because $\epsilon^p$ or $\epsilon^{|1-p|}$ then changes slowly.

\subsection{Reference-subtracted angular profiles}
Set $x_0=1/3$. Subtracting $\Phi(x_0)$ removes the common divergence and matching constant. The fixed-throat and $0<p<1$ limiting profile is
\begin{equation}
 F(x)=-\ln\!\left[\frac{27}{4}x(1-x)^2\right].
 \label{eq:fixed_normalized}
\end{equation}
It has a unique minimum at $x_0$, because $F'=(3x-1)/[x(1-x)]$ and $F''=1/x^2+2/(1-x)^2>0$. Along the power-law paths,
\begin{equation}
 \Phi(x)-\Phi(x_0)\longrightarrow
 \begin{cases}
 F(x),&0<p<1,\\
 G_\kappa(x)-G_\kappa(x_0),&p=1,\\
 G_0(x)-G_0(x_0),&p>1.
 \end{cases}\label{eq:path_profiles}
\end{equation}
The reference sets the zero, not the minimum of a general crossover profile. For $\kappa=1$ the latter lies near $x=0.506395$; the $K=0$ minimum is at $(\sqrt5-1)/2\simeq0.618034$. The limiting shapes are connected by $K\to\infty$ and $K\to0$ in Eq.~\eqref{eq:crossover_shape}, but remain distinct for the three path sectors.

Figure~\ref{fig:profiles} resolves information hidden by the common logarithmic growth. In the left panel, the skewed numerical profiles approach $F$. In the right panel, the green and purple families have distinct limiting minima away from the reference ray. Their solid curves nearly overlap the corresponding black dashed limits, reflecting convergence within each family. The sampled maximum errors at $\epsilon=10^{-6}$ are approximately $3.64\times10^{-3}$, $1.35\times10^{-6}$ and $1.91\times10^{-3}$ radians for $p=1/2,1,3/2$. Using the finite $K$ composite profile instead reduces these sampled discrepancies to below $1.6\times10^{-6}$ radians in all three cases. The distinction reflects convergence of $K$, in addition to the local geometric approximation.

The plotted integrals use inverse areal radius at 55-digit precision, with a factorized radial cubic and a squared-distance substitution at the inner turning root. Independent areal-radius quadratures checked seven rays across fixed-throat and power-law approach paths, including $\kappa=0.5,2,5$, with absolute angular differences below $4\times10^{-44}$ radians at high arithmetic precision. Two controls at 75 digits tested precision convergence. These finite numerical checks complement the compact-channel remainder estimate; critical endpoints require their own scaling.

\subsection{Aperture on the observer's sky}
For a static observer at end-1 areal radius $r_o\gg b_c$, the local angle satisfies $\sin\theta=b\sqrt{f_1(r_o)}/r_o$ \cite{PerlickTsupko2022}. Let $\theta_\pm$ correspond to $b_\pm$. The solid angle of the reflected annulus is
\begin{equation}
 \begin{split}
 \Delta\Omega&=2\pi(\cos\theta_--\cos\theta_+)\\
 &=\frac{\pi f_1(r_o)b_c^2}{r_o^2}\,D
 \left[1+O\!\left(\frac{b_c^2}{r_o^2}\right)\right].
 \end{split}\label{eq:aperture}
\end{equation}
Here $D$ is the exact squared-impact gap of Eq.~\eqref{eq:D}, and $\Delta\Omega$ is dimensionless. The angular aperture has the width exponent $q$, even while the retained rays acquire increasing winding. Its geometric suppression does not fix the received flux: that additionally weights the admitted disk intersections by an emission prescription \cite{Gralla2019,GLM2020}.

\begin{figure*}[!tb]
\begin{minipage}{\textwidth}
\centering\includegraphics[width=\textwidth]{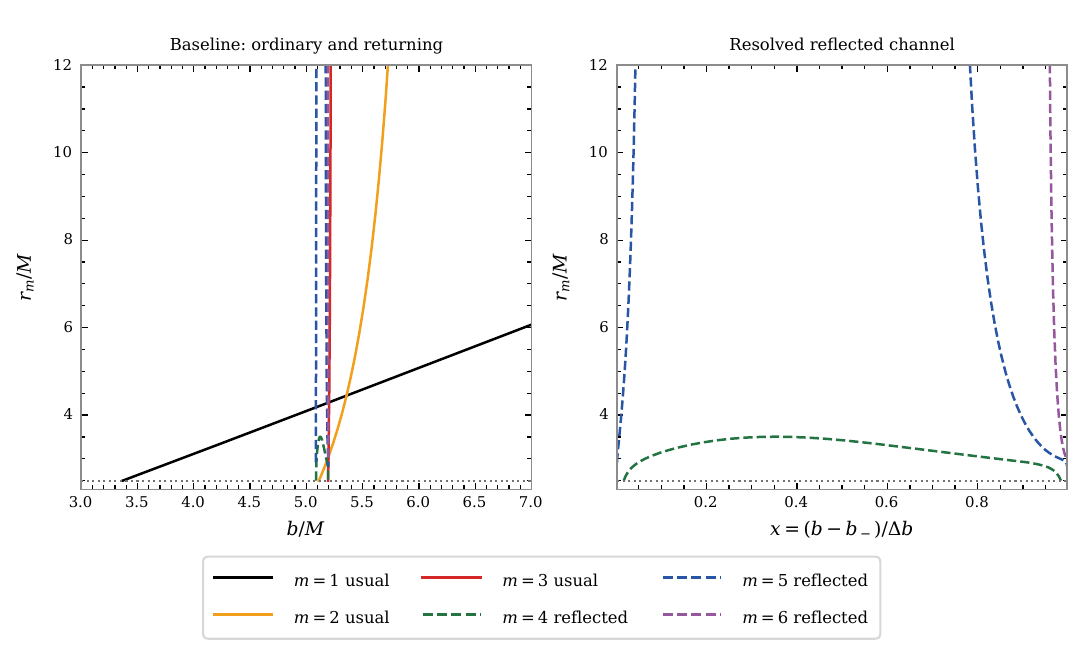}
\caption{Geometrical disk-transfer functions for the baseline $a/M=2.5$, $\epsilon=0.02$, with a thin transparent disk in end 1 and an axial observer. Left: an impact-space overview, showing the usual incoming/scattered branches with $m=1,2,3$ as solid curves and selected final-return branches with $m=4,5,6$ as dashed curves. Right: those reflected branches resolved in the channel coordinate over $0.002\leq x\leq0.998$. Color identifies $m$ in both panels. The index is the angular label of Eq.~\eqref{eq:disk_phase}, with end-2 passages omitted from the disk transfer. The dotted level is the disk edge $r=a$. Radii above $12M$ leave the display window and are not capped numerically; higher angular indices are not exhausted by the displayed selection.}
\label{fig:disk_transfer_baseline}
\end{minipage}
\end{figure*}

\section{Geometrical transfer to a thin disk}\label{sec:disk_transfer}
Place a transparent, geometrically thin test disk in end 1, perpendicular to an asymptotic observer's axis and extending over $a\leq r<\infty$. It is an intersection surface; gas velocity, emissivity and an orbital inner edge are not imposed. With angular origin on the observer's axis in each ray plane, successive passages through the aligned disk plane occur at
\begin{equation}
 \phi_m=\frac{(2m-1)\pi}{2},\qquad m=1,2,\ldots.
 \label{eq:disk_phase}
\end{equation}
The positive integer $m$ labels global angular plane crossings. Only passages in end 1 at $r\geq a$ define disk-transfer radii $r_m(b)$; end-2 passages do not emit in this unilateral model. Allowed labels can therefore have gaps and need not count emitting intersections consecutively. The underlying return-stage selection and disk-transfer mechanism were established in Ref.~\cite{Peng2021}; the issue here is their behavior as the reflected channel closes.

\begin{figure*}[!tb]
\begin{minipage}{\textwidth}
\centering\includegraphics[width=\textwidth]{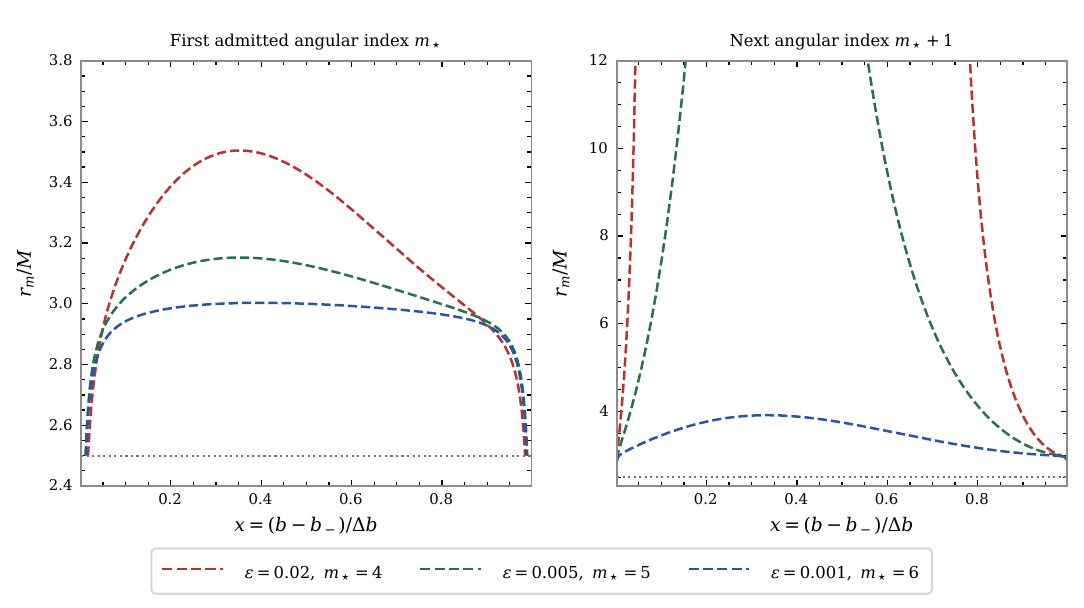}
\caption{Channel-resolved transfer for three asymmetries at fixed $a/M=2.5$. Color denotes $\epsilon$ and the legend specifies its first admitted angular index $m_\star$. Left: the branch with that index. Right: the next angular branch, $m_\star+1$. All curves are final-return intersections in end 1, and $0.002\leq x\leq0.998$ is common to the panels. The vertical scales differ: the left panel resolves $2.4\leq r_m/M\leq3.8$, while the right extends to $12$. The dotted level marks the disk edge. An interval with no admitted return crossing is distinguished in the text from a connected branch passing above the radius window.}
\label{fig:disk_transfer_asymmetry}
\end{minipage}
\end{figure*}

\subsection{Return support and radial inversion}
Define the partial end-1 angle
\begin{equation}
 \mathcal A(r;b)=\int_r^\infty
 \frac{b\,\dd r'}{r'^2\sqrt{1-b^2U_1(r')}}.
 \label{eq:disk_partial_angle}
\end{equation}
This dimensionless function decreases with radius and obeys $\mathcal A(a;b)=\varphi_1$. A crossing on the final end-1 return segment exists precisely when its phase lies between throat re-entry and escape:
\begin{equation}
 n_2(b)<\frac{2m-1}{4}<n_3(b).
 \label{eq:disk_return_support}
\end{equation}
This is the general angular-index form of the return inequalities used in Ref.~\cite{Peng2021}. Equality on the left places the crossing at the inner disk edge; equality on the right sends it to infinity. The admitted radius satisfies
\begin{equation}
 \mathcal A\!\left(r_m^{\rm ret};b\right)=\Phi(b)-\phi_m.
 \label{eq:disk_return_radius}
\end{equation}
The superscript ret denotes the final return segment, whose phase at $r$ is $\Phi-\mathcal A(r;b)$. Under Eq.~\eqref{eq:disk_return_support}, the right-hand side lies strictly between zero and $\varphi_1$, so the monotonic inversion gives exactly one radius in $a<r<\infty$.

The overview in Fig.~\ref{fig:disk_transfer_baseline} separates ordinary low-index intersections from the narrow reflected contribution. In the baseline, the first admitted final-return label in the adopted channel window is $m=4$; stopping at $m=3$ would omit this family. This label is not automatically a fourth image, because previous global plane crossings can lie outside the emitting exterior.

The resolved $m=4$ branch meets the disk edge at both support boundaries. The next label, $m=5$, instead has disconnected support, with interior endpoints near $x=0.0700$ and $x=0.7157$ at which its radius diverges. The $m=6$ branch lies still closer to the channel edges, and parts exceed the displayed $12M$ radius range. These mechanisms are distinct: Eq.~\eqref{eq:disk_return_support} determines whether the intersection exists; the ordinate window determines whether it is visible in the plot.

\subsection{Migration of the first admitted angular label}
Let $m_\star$ be the smallest admitted angular index anywhere in $0.002\leq x\leq0.998$. At fixed $a/M=2.5$, minimizing the return-start angle and checking both support inequalities gives $m_\star=4,5,6$ for $\epsilon=0.02,0.005,0.001$. The minimum values of $n_2$ are approximately $1.31477$, $1.75486$ and $2.26668$. The aperture therefore contracts while its first available return crossing shifts to a later global label.

Figure~\ref{fig:disk_transfer_asymmetry} compares branches selected by admission. Their first available intersections remain near the throat and photon-sphere scale, but retain different radial profiles after the impact interval has been normalized. Since $m_\star$ changes between curves, the ordering of their sampled peak radii is not a fixed-index limit or a general monotonic law in $\epsilon$. The differences retain the fractional phase left after the increasing winding is compared with the discrete crossing phases.

The next-index panel distinguishes the genuine split red $m=5$ support from the connected green $m=6$ branch whose central radii exceed $12M$. Its apparent gap is only a display effect; the blue $m=7$ branch remains within the window. This distinction survives any common horizontal normalization because it is determined by the phase inequalities and radial inversion.

Support boundaries were bracketed in angular phase and the radial inversion used an adaptively checked curve mesh. Independent areal-radius quadratures tested nine reflected and six ordinary intersections; the largest reflected phase residual was below $1.5\times10^{-9}$ radians and the largest independently inverted radius difference below $1.1\times10^{-8}M$. These tests concern geometrical transfer in the sampled domain. Turning the annular aperture into a flux requires weights for all emitting crossings and control of their radial support.

\FloatBarrier
\section{Conclusions}\label{sec:conclusions}
The loss of the reflected family near symmetry is governed by the competition between the throat location and the closing critical-impact gap. Both bulk photon-sphere maxima remain nondegenerate. Their local instability alone does not determine how much of a complete ray can access each near-critical region.

A normal-coordinate construction gives a composite accumulated angle with a controlled remainder on compact interior intervals of the reflected channel. It remains valid when the ratio of throat distance to mass contrast tends to zero or infinity. Power-law approaches consequently interpolate between width exponents one and two, while the coefficient of logarithmic winding decreases from three to two and then saturates. The fixed-throat amplitude remains distinct from the small-distance expansion. Reference subtraction exposes three connected sectors of finite angular profiles rather than removing their geometrical differences.

The observer's annular solid-angle aperture inherits the width exponent. For a transparent unilateral test disk, the established return-phase condition then distinguishes available intersections from global plane passages in the opposite exterior. In the sampled fixed-throat sequence, the first admitted return label increases as the aperture contracts. Disconnected support and connected branches outside a finite radius window remain physically different cases.

The composite expansion supplies a starting point for endpoint scalings and for radiative models with specified source motion and radial support. Those extensions can determine how the increasingly wound paths are weighted within the vanishing angular aperture.

\section*{Acknowledgments}
Edson Otoniel acknowledges support from Funda\c{c}\~ao Cearense de Apoio ao Desenvolvimento Cient\'ifico e Tecnol\'ogico through grant BP6-0241-00335.01.00/25.

\bibliographystyle{elsarticle-num}
\bibliography{ref}
\flushcolsend
\end{document}